# Conformal oxide coating of Carbon Nanotubes**

*S. Kawasaki, G. Catalan, H. Fan, M. M. Saad, J. M. Gregg, M. A. Correa-Duarte, J. Rybczynski, F. D. Morrison, T. Tatsuta, O. Tsuji,* and *J. F. Scott**

The broad use of ferroelectric oxide films as capacitors in dynamic random access memories (DRAMs) has been developed, and ferroelectric random access memories (FeRAMs) are now also important commercial products as non-volatile memories.[1] Considerable ferroelectrics research has been focused on growth in integration density of FeRAMs, with the latest commercial device in production an 8 Mb PZT memory from Texas Instruments. However, all FeRAMs are still planar-stacked devices, even though the International FeRAM Roadmap[2] suggests that FeRAM structures must become three-dimensional (3D) by 2010 to accommodate requisite storage density.

Recently hybrid approaches are being tried, whereby structures using nanowires or nanotubes are grown, rather than patterned and etched[3-5]; and this is a possible solution to maintain the growth in integration density. For ultra-high integration density DRAMs, metal-dielectric insulator/carbon-nanotube/metal layers on multi-wall carbon nanotubes (MWCNTs) normal to substrates have already been fabricated with plasma enhanced chemical vapor deposition (PECVD).[6] We believe that the combination of ferroelectric thin films and MWCNTs can be a great candidate for reducing capacitor size in integrated FeRAM circuits. However, ferroelectric-MWCNTs structures have never been fabricated because MWCNTs would not survive the deposition process for crystalline ferroelectric thin films; the latter require crystallization at high temperatures with oxygen ambient, an environment in which the CNTs would literally burn up.

On the other hand, the liquid source misted chemical deposition (LSMCD) technique was developed[7] as a novel deposition technique for multi component oxide thin films and has been used to fabricate various kinds of thin films,[8-11] such materials as ferroelectrics and high-k dielectrics. One of the primary advantages of LSMCD technique is that it uses chemical solution deposition (CSD) precursors that are not subjected to high temperature during the deposition process, and the crystallization process with high temperature in oxygen ambient is complete in a matter of minutes, so that the thermal budget is quite small. In addition, this deposition scheme uses ordinary sol-gel precursors but eliminates spin coating, so that three-dimensional nano-structures can be produced, including nanotubes and nano-wires. The approach to achieve ferroelectric-MWCNT structures with LSMCD techniques allows the possibility of overcoming the combustion of MWCNTs during the deposition for crystalline ferroelectric thin films.

Also, the excellent step coverage performance of LSMCD process is essential to achieve conformal coating of ferroelectric films on MWCNTs. Morrison *et al*[12] successfully demonstrated conformal coating on walls of porous Si holes with high aspect ratio of 100 or more via the LSMCD technique, and individual free-standing regular arrays of $SrBi_2Ta_2O_9$ ferroelectric nanotubes of diameter 400 nm to a few microns were obtained. Many groups have also fabricated nano/micro tubes with related infiltration techniques[13-16] that gave wetting and complete filling of the pore wall of porous templates by source precursors; evaporating the precursors thereby yielded nano- or micro-tube structures. However, such infiltration-derived nano/micro tubes are unsuitable for convex structures, such as vertical free-standing nanowires.

In the present work we have modified the conventional LSMCD technique to achieve conformal coating of 3D nanostructures by applying low temperature heating to both the mist and the template during deposition. As a specific embodiment of the process, we give details here of the deposition of ferroelectric $Pb(Zr,Ti)O_3$ (PZT) on MWCNT arrays.

The MWCNTs arrays used in this work were grown by PECVD on Ni catalyst dots on Si wafer. $NH_3$ was used as a plasma enhancer and growth promoter. After exposing the substrate under plasma, $C_2H_2$ was introduced into the chamber to trigger the MWCNTs growth at 5-10 Torr. A detailed description of MWCNTs growth through this process can be found elsewhere.[17] PZT films were deposited onto the MWCNT arrays. The metalorganic decomposition-type PZT solution (17 wt%) synthesized by Kojyundo Chemical Laboratory Co., Ltd. (Japan) was used as the starting material. The composition of the solution in the ratio of Pb/Zr/Ti was 1.1/0.4/0.6. The concentration of metal components in the precursor solution was adjusted to 4 wt% by diluting PZT starting material with methyl ethyl ketone. The precursor solution was converted to mist by an atomizer. A pipe located between an atomizer and nozzle is heated at 140/300ºC, and heated mist sprayed out vertically on the top of MWCNT arrays at atmospheric pressure. This MWCNT template was also heated at 120/260°C during the mist deposition. The deposited PZT precursor layer was transformed into amorphous oxide layer by pyrolysis at 300°C for 3 min and subsequently crystallized with tube furnace at 650°C for 5 min under oxygen flow of 2 L/min.

[*]   Prof. J. F. Scott, Dr. S. Kawasaki, Dr, G. Catalan, Dr. H. Fan
Earth Sciences Department, University of Cambridge, Cambridge CB2 3EQ, UK
E-mail: jsco99@esc.cam.au.uk; skaw05@esc.cam.ac.uk
Dr. J. M. Gregg, M. M. Saad
Department of Pure and Applied Physics, Queen's University of Belfast BT7 1NN, UK
Dr. M. A. Correa-Duarte
Departamento de Quimica Fisica, Universidad de Vigo, Spain
Dr. J. Rybczynski
Department of Physics, Boston College, MA 02467, USA
Dr. F. D. Morrison
School of Chemistry, University of St Andrews, North Haugh, St Andrews KY16 9ST, UK
O. Tsuji, Dr. T. Tatsuta
SAMCO Inc., 36 Waraya-cho, Takeda, Fushimi-ku, Kyoto 612-8443, Japan

[**] The authors wish to acknowledge the Kojyundo Chemical Laboratory Co., Ltd. for preparing the PZT liquid source precursor. This work has been partially supported by the Spanish Xunta de Galicia under grant No. PGIDIT03TMT30101PR. GC acknowledges financial support of a Marie Curie Fellowship.

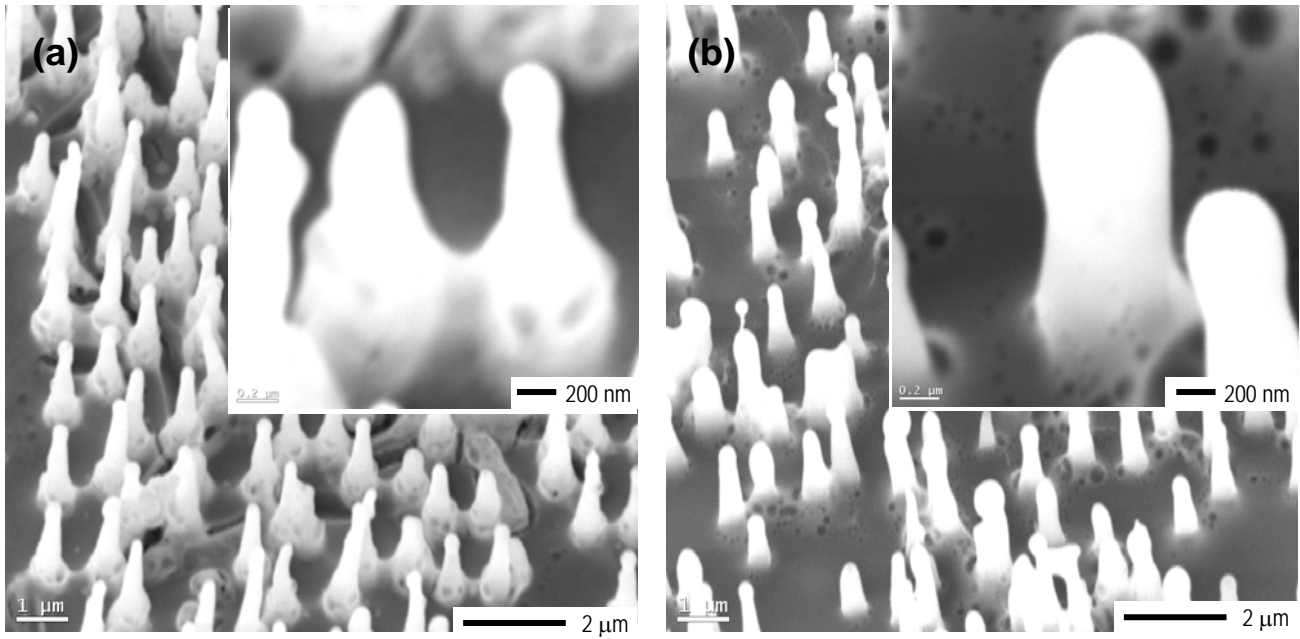

**Figure 1.** SEM images of PZT films deposited onto MWCNTs at template temperature of 120°C with mist heated at 140°C (a), and template temperature of 260°C with mist heated at 300°C (b).

Figure 1 presents scanning electron microscope (SEM) images of PZT film deposited on MWCNTs after crystallization. The PZT film in Figure 1a was deposited at template temperature of 120°C with mist heated at 140°C. The so-called Stone-Walls (SW) defects, which are pairs of adjacent five and seven member rings and are commonly present on the CNT walls, are generally accepted as active defect sites, helping here the deposition of the PZT. Usually CNTs can be oxidized to $CO_2$ gas in an oxygen atmosphere at temperatures above 400°C because they consist of carbon atoms connected to each other by covalent bonds. However, and despite of this implementation of high temperature annealing under oxygen ambient needed to crystallize PZT film, the survival of MWCNTs can be explained taking into account the amorphous oxide layer, pyrolyzed at 300°C for 3 min. This protecting shell encapsulating the CNTs avoids their complete oxidation, in a similar way as when encapsulated in ceramic materials, protecting the CNTs at temperatures up to 700°C.[18] However, the higher resolution image in the inset of Figure 1a clearly indicates that most of the coating occurs near the bottom of the MWCNTs. This result suggests that the majority of the deposited precursor slides down the sides of the tubes, as the mist is still too "wet". Figure 1b shows the surface images of PZT coated MWCNTs with higher deposition temperature of template and mist, 260°C and 300°C, respectively. In contrast with lowered temperature deposition, the film morphology is completely continuous and the film coverage is conformal around the tip and the bottom of the MWCNTs in the inset of Figure 1b. Moreover, as can be seen in the lower resolution image in Figure 1b, this perfect conformality was achieved over a large area.

To evaluate the step coverage on MWCNTs, we have analyzed the diameter of coated MWCNT from SEM images with 10 points in each of the 5 positions as described in the schematic of Figure 2a. The average diameter of each position measured at several places on the template is shown in Figure 2b. The improvement of step coverage in coating onto MWCNTs at higher temperatures was confirmed. This is due to the immediate drying out of deposited mist before it slides down to the bottom of MWCNT. Figure 3a shows cross-sectional transmission electron microscope (TEM) image of PZT coated MWCNTs deposited at template temperature of 260°C. In Figure 3b, energy dispersive X-ray spectroscopy (EDS) and mapping shows that deposited film contains Pb and oxygen, although the Zr and Ti concentrations are lower than stoichiometric. No crystallographic identification of PZT (e.g., from TEM) has yet been made; nor have attempts been made to distinguish pyrochlore from perovskite phases. Related studies in our group suggest that the product contains both PZT and PbO.[19] However, the MWCNT appear to have siphoned up some of the Si from substrate. Since the reacted MWCNTs were then not very good conductors, ferroelectric and piezoelectric properties cannot yet be characterized. By using metallic substrates we will presently try to avoid this migration during the crystallization process.

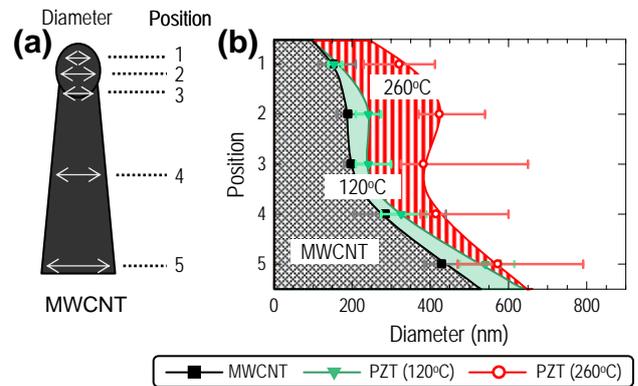

**Figure 2.** Schematic illustration showing the PZT-coated MWCNT diameter position assumed from SEM images (a), and average diameter of each position measured at several places on template (b).

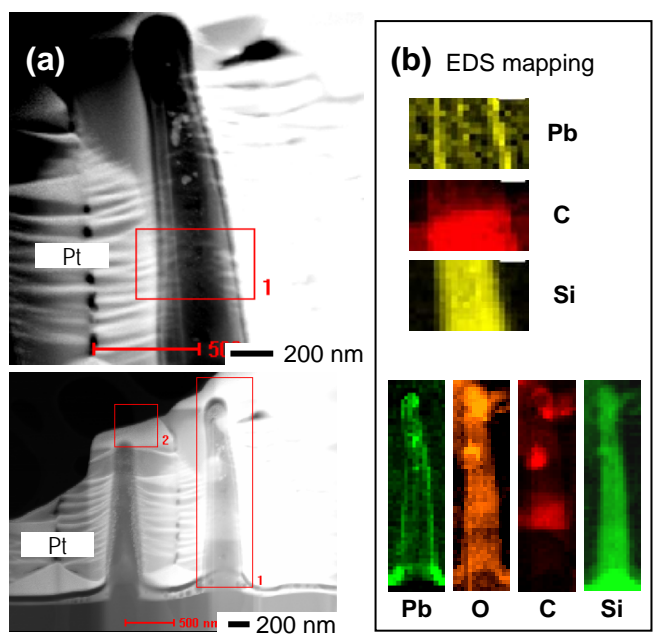

**Figure 3.** Cross-section TEM images of PZT coated MWCNTs (a), and EDS and mapping of Pb, O, C and Si (b).

In conclusion, we have developed a technique to control mist penetrability and fluidity to ensure good step coverage of carbon nanotubes while preventing their combustion. We have used this technique to conformally deposit PZT as a first step towards the eventual integration in future 3D FeRAM nanostructures. We recognize that MWCNTs need to be grown on substrate with conductive layer, such as Nb, Pt, etc. to preserve MWCNTs as high conductive electrodes, thus avoiding Si contamination in the MWCNTs from migrated material from the substrate. This, in turn, will enable the verification of ferroelectric switching. The nanoscale capacitor structures of ferroelectric PZT film with MWCNTs electrodes have the potential to achieve ultrahigh integration density FeRAMs, thus replacing the Si pillar structure in future.

[1]  J. F. Scott, *Science* **2007**, 315, 954.
[2]  International Technology Roadmap for Semiconductors 2006 Update, Front End Process.
[3]  A. M. Morales, C. M. Lieber, *Science* **1998**, 279, 208.
[4]  J. Liu, X. Li, L. Dai, *Adv. Mater.* **2006**, 18, 1740.
[5]  P. R. Evans, X. Zhu, P. Baxter, M. McMillen, J. McPhillips, F. D. Morrison, J. F. Scott, R. J. Pollard, R. Bowman, J. M. Gregg, *Nano Lett.* **2007**, 7, 1134.
[6]  J. E. Jang, S. N. Cha, Y. Choi, G. A. J. Amaratunga, D. J. Kang, D. G. Hasko, J. E. Jung, J. M. Kim, *Appl. Phys. Lett.* **2005**, 87, 263103.
[7]  L. D. McMillan, C. A. P. Araujo, T. Roberts, J. Cuchiaro, M. C. Scott, J. F. Scott, *Integr. Ferroelectr.* **1992**, 2, 351.
[8]  S. Kawasaki, S. Motoyama, T. Tatsuta, O. Tsuji, S. Okamura, T. Shiosaki, *Jpn. J. Appl. Phys.* **2004**, 43, 6562.
[9]  K. Woong, M. K. Jeon, K. S. Oh, T. S. Kim, Y. S. Kim, S. I. Woo, *PNAS* **2007**, 104,1134.
[10] J. Jiang, O. O. Awadelkarim, D. O. Lee, P. Roman, J. Ruzyllo, *Solid State Electr.* **2002**, 46, 1991.
[11] S. Kawasaki, S. Motoyama, T. Tatsuta, O. Tsuji, T. Shiosaki, *Integr. Ferroelectr.* **2003**, 53,287.
[12] F. D. Morrison, L. Ramsay, J. F. Scott, *J. Phys.: Condens. Matter.* **2003**, 15, L527.
[13] M. Steinhart, J. H. Wendorff, A. Greiner, R. B. Wehrspohn, K. Nielsch, J. Schilling, J. Choi, U. Gosel, *Science* **2002**, 296, 1997.
[14] Y. Luo, L. Szafraniak, N. D. Zakharov, V. Nagarajan, M. Steinhart, R. B. Wehrspohn, J. H. Wendorff, R. Ramesh, M. Alexe, *Appl. Phys. Lett.* **2003**, 83, 440.
[15] S. S. N. Bharadwaja, M. Olszta, S. Trolier-McKinstry, X. Li, T. S. Mayer, F. Roozeboom, *J. Am. Ceram. Soc.* **2006**, 89, 2695.
[16] X. Li, F. Cheng, B. Guo, J. Chen, *J. Phys. Chem. B* **2005**, 109, 14017.
[17] Y. Wang, J. Rybczynski, D. Z. Wang, K. Kempa, Z. F. Ren, W. Z. Li, B. Kimball, *Appl. Phys. Lett.* **2004**, 85, 4741.
[18] A. H. Cannon, A C. Allen, S. Graham, W. P King, *J. Micromech, Microeng.* **2006**, 16, 2554.
[19] X. Lou, M. Zhang, S. A. T. Redfern, J. F. Scott, *Phys. Rev. Lett.* **2007**, 97, 177601.